\documentclass[11pt, A4]{article}
 
\usepackage{graphicx,subfigure}
\usepackage{epstopdf}
\usepackage{color}
\usepackage{xcolor}
\usepackage{epsfig,amssymb,amsmath}
\usepackage{amsthm}
\usepackage{empheq}
\usepackage{mathrsfs}

\usepackage{amsfonts}
\usepackage{amsthm}

\definecolor{GC}{rgb}{0,0.0,0.65}
\usepackage[
      colorlinks=true,
      linkcolor=blue,
      urlcolor=blue,
      filecolor=blue,
      citecolor=red,
      pdfstartview=FitV,
      pdftitle={},
      pdfauthor={},
      pdfsubject={},
      pdfkeywords={},
      pdfpagemode=None,
      bookmarksopen=true
]{hyperref}

\usepackage{epsfig}

\usepackage{hyperref}

\def\cC{\mathcal{C}}
\def\cD{\mathcal{D}}

\def\cM{\mathcal{M}}

\def\cR{\mathcal{R}}

\def\d{\partial}

\def\beq{\begin{eqnarray}}
\def\eeq{\end{eqnarray}}

\def\a{\alpha}
\def\b{\beta}
\def\g{\gamma}
\def\d{\delta}

\def\o{\omega}

\def\s{\sigma}

\def\o{\omega}

\def\be{\begin{equation}}
\def\ee{\end{equation}}
\def\bea{\begin{eqnarray}}
\def\eea{\end{eqnarray}}

\newcommand{\rom}[1]{\mathrm{#1}}

\def\cC{\mathcal{C}}
\def\cD{\mathcal{D}}

\def\cM{\mathcal{M}}

\def\cR{\mathcal{R}}

\def\nn{\nonumber}

\def\curl{\mathrm{curl\,}}

\newcommand*\widefbox[1]{\fbox{\hspace{1em}#1\hspace{1em}}}

\begin{document}
\pagestyle{myheadings}
\markboth{\textsc{\small }}{%
  \textsc{\small Supertranslations and Holographic Stress Tensor}}
  \addtolength{\headsep}{4pt}

\begin{flushright}
\texttt{AEI-2011-100}\\
\end{flushright}
\vspace{1cm}

\begin{centering}

  \vspace{0cm}

  \textbf{\Large{Supertranslations and Holographic Stress Tensor}}

  \vspace{0.8cm}

  {\large Amitabh Virmani}

  \vspace{0.5cm}

\begin{minipage}{.9\textwidth}\small  \begin{center}

Max-Planck-Institut f\"ur Gravitationsphysik (Albert-Einstein-Institut) \\
Am M\"uhlenberg 1, 
D-14476 Golm, Germany \\
{\tt virmani@aei.mpg.de}
\\ $ \, $ \\

\end{center}
\end{minipage}

\end{centering}

\vspace{1cm}

\begin{abstract}
It is well known in the context of four dimensional asymptotically flat
spacetimes that the leading order boundary metric must be conformal to unit de
Sitter metric when hyperbolic cutoffs are used. This
situation is very different from asymptotically AdS settings where one is
allowed to choose an arbitrary boundary metric. The closest one can come to changing the boundary metric in the asymptotically
flat context,
while maintaining the group of asymptotic symmetries to be Poincar\'e,
 is to change the so-called `supertranslation frame' $\o$.
 The most studied choice corresponds to taking $\omega = 0.$ 
 In this paper we study consequences of making alternative choices. We perform
 this analysis in the covariant phase space approach as well as in the
 holographic renormalization approach. We show that all choices for $\omega$ are
 allowed in the sense that the covariant phase space is well defined
 irrespective of how we choose to fix supertranslations. The on-shell action and
 the leading order boundary stress tensor are insensitive to the
 supertranslation frame. The next to leading order boundary stress tensor
 depends on the supertranslation frame but only in a way that the transformation
 of angular momentum under translations continues to hold as in special
relativity.
\end{abstract}
\vfill

\thispagestyle{empty} \newpage

\tableofcontents

\newpage

\setcounter{equation}{0}
\section{Introduction}
The development of gauge/gravity dualities has revolutionized string theoretic
 investigations of quantum gravity. These dualities relate string theories in
 higher dimensions to certain quantum theories in lower dimensions on fixed
 metric backgrounds. Consequently, they provide us with a framework where one
 can address deep puzzles of quantum gravity concerning black holes, singularities and the
 like, by performing calculations in lower dimensional non-gravitating settings.
 The best understood case occurs in AdS/CFT \cite{M} where string theory on AdS
 space space is dual to certain gauge theory on the boundary of AdS space.
 Similar dualities are not known for spacetimes of the most physical interest,
 such as cosmologies or flat Minkowski space. It is clearly of interest to
 investigate if holographic dualities exist in these settings.

A well appreciated point is that although the
most well understood cases of these dualities involve AdS spaces and local CFTs, neither AdS nor CFTs are
fundamental to these dualities. It is often speculated that every possible
boundary condition that defines a bulk string theory is dual to a
non-gravitational theory through the AdS/CFT correspondence. In the
case of AdS spaces, the correspondence arises from the way in which the AdS boundary parametrizes the space of 
possible boundary conditions on bulk fields \cite{Wi, GKP}. In fact, the
richness of AdS/CFT precisely comes from the fact that in an asymptotically AdS settings one
is allowed to choose a variety of boundary conditions. In particular, one is
allowed to choose an arbitrary boundary metric \cite{Sk1, BK1, BK2, Sk2, Sk3}.

The corresponding situation in the asymptotically flat setting is very
different. This is one of the reason why the the subject of holographic duality
for flat spacetime has resisted developments over all these years. For
example, for the four dimensional asymptotically flat spacetimes the leading
order boundary metric must be conformal to unit de Sitter metric when hyperbolic cutoffs are used to define asymptotically flat
spacetimes \cite{BS}. The structure of the asymptotic equations is such
that there is absolutely no freedom in the choice of the leading order boundary metric.
One is led to ask how much freedom one has at the next to leading order. Even
there the freedom is quite limited. As we discuss in detail in this paper, the
closest one can come to changing the boundary metric in the asymptotically flat context, while 
maintaining asymptotic symmetries to be Poincar\'e, 
is to change the so-called `supertranslation frame' $\o$. The information about
the supertranslation frame enters in the asymptotic expansion is a very specific
way. In this paper we study what it precisely means to change the
supertranslation frame, and the consequences this brings to the construction of
the boundary stress tensor.

We perform this analysis in the covariant phase space approach of \cite{ABR,
LW, W, IW, WZ} as well as in the holographic renormalization approach of Mann
and Marolf \cite{MM, MMV, MMMV}. The key results of the present paper are as
follows. First, we show that all choices for the supertranslation frame are allowed. More precisely, that covariant phase
space is well defined irrespective of how we choose to fix supertranslations.
Second, we show that the on-shell action and the leading order boundary stress
tensor are insensitive to the supertranslation frame. Third, we show that the
next to leading order boundary stress tensor depends on the supertranslation frame but only in a way 
that the transformation of angular momentum under translations continues to hold as 
in special relativity.

We will elaborate on these points momentarily, for now let us step back a bit
and recall certain basic facts about supertranslations. It turns out that the
issue of supertranslations is much related to the general notion of angular momentum. Recall
that in special relativity the notion of angular momentum
is origin dependent. This origin dependence arises because there is a
four-parameter family of Lorentz subgroups in the Poincar\'e group. None of
these subgroups is preferred over any other, so the origin dependence is
inevitable. The structure of the Lie algebra of the Poincar\'e group then tells
us the transformation property of the angular momentum under the action of
translations. The resulting notion matches with our intuitive understanding of
angular momentum.

For asymptotically flat spacetimes at spatial infinity, the asymptotic
symmetries form  an infinite dimensional group---the so-called spatial
infinity (SPI) group \cite{Ashtekar:1978zz, Ashtekar:1991vb}. The SPI group is
similar to the Poincar\'e group, except that the four translations are replaced
by an infinite number of angle dependent translations---the so called
supertranslations. The group structure of the SPI group is that of a
semi-direct product of the supertranslation group with the Lorentz group. The
supertranslation group is the infinite-dimensional additive group of smooth
functions on the unit hyperboloid. The semi-direct product structure is as
follows: if $(\alpha, \xi^a)$ and $(\beta, \eta^a)$ are two elements of the Lie algebra of
the SPI group, with $\alpha$ and $\beta$ arbitrary smooth functions on the
hyperboloid and $\xi^a$ and $\eta^a$ exact Killing vectors of the unit
hyperboloid, then the SPI Lie bracket is \cite{Ashtekar:1978zz, Ashtekar:1991vb}
 \be
[(\alpha, \xi^a),(\beta, \eta^a)] = (\pounds_\xi \beta - \pounds_\eta \alpha,
[\xi, \eta]^a).
\ee
It follows that the SPI group admits an infinite class of Lorentz
subgroups.  None of these subgroups is preferred over the
others.
Therefore, a naive approach to defining angular momentum suffers from the so-called supertranslation
ambiguities: origin dependence of angular momentum where the origin lies in an
infinite dimensional space. We clearly need an additional structure at spatial
infinity that can reduce the SPI group to the Poincar\'e group. These points
were very well emphasized in \cite{Ashtekar:1978zz,AM2}. The main
purpose of this paper is to systematically study the freedom we have in the Poincar\'e reduction of the SPI
group.

The plan of the rest of the paper is as follows. In section
\ref{sec:review} we begin with various definitions and provide a brief review of
the counterterm construction of \cite{MM}. In this section we also review the
boundary conditions of Ashtekar, Bombelli, and Reula (ABR) \cite{ABR}, and
present our supertranslated generalization of the ABR boundary conditions. The
main point of this section is to show that the covariant phase space is well
defined for our supertranslated boundary conditions. 

In section \ref{sec:expansion} we perform systematic expansion of the equations
of motion and discuss Beig's integrability conditions \cite{B, CDV}. We find
that despite the fact that we have an arbitrary function $\omega$ in
our asymptotic expansion, the integrability conditions do not change. As
in \cite{B, CDV}, the integrability conditions require Lorentz charges
constructed using
\be 
\curl [ 4
\epsilon_{cd(a} \s^c \s^{d}_{b)}]
\ee 
to be zero. In this paper we choose, following ABR \cite{ABR}, mass
aspect $\sigma$ to be a symmetric function on the hyperboloid. As a result, the 
integrability conditions are automatically satisfied \cite{MMMV, CDV}. In this 
section we also  perform a systematic expansion of the
Mann-Marolf counterterm.

Section \ref{sec:BST} is devoted to the study of the renormalized on-shell
action and the expansion of the boundary stress tensor.  We show that the on-shell action
and the leading order boundary stress tensor are insensitive to the
supertranslation frame.  In section \ref{sec:BSTprop} properties of the boundary
stress tensor are studied in further detail.  Our boundary
stress tensor satisfies all the expected properties: (i) it is conserved a la
Brown-York \cite{BY}, (ii) it reduces to the previous expression of \cite{MMMV}
when either (a) $\omega =0$ or (b) when $\omega$ represents a
translation---i.e.,
when it is not a non-trivial supertranslation, and  (iii) the next to leading
order boundary stress tensor transforms under translations in an expected way.
 Finally, in section \ref{sec:conclusions} we end with our conclusions and
 possible future directions. Certain technical and computational details are
 relegated to two appendices. In appendix \ref{app:EOM} asymptotic expansion of
 the equations of motion is presented. In appendix \ref{app:BST} certain details
on the asymptotic expansion of the boundary stress tensor are presented.

 As a last comment in this section, we wish to emphasize an
 important point. In the study of asymptotic structure of spacetimes,
 the notions one introduces and the boundary conditions one chooses are to some
 extent arbitrary; their justification lies in the perspective they bring. 
 Already there are variety of methods known to analyze asymptotic flatness and
 construct conserved quantities \cite{WZ, Ashtekar:1978zz, ADM1,  RT, Geroch, 
 AshRev, AD,  BB, Sorkin}. These different approaches offer different
 perspectives.
 All these methods ultimately lead to similar/equivalent results.
 For example, 
 the framework presented in \cite{Ashtekar:1978zz} defines unambiguously a
 useful notion of angular momentum at spatial infinity, and allows us to relate
such construction to the analogous conserved quantities at null infinity.
The boundary conditions of \cite{Ashtekar:1978zz} are further strengthened in
\cite{ABR} by demanding mass aspect $\s$ to be symmetric. These strengthened
boundary conditions offer a new perspective:
they lead to a well defined covariant phase space. The study presented below should
be taken in this spirit. Our motivation is a combination of the ideas: (i) we wish to have an unambiguous and useful notion
of angular momentum, (ii) we wish to have a well defined phase space precisely
in the sense of \cite{ABR}, and finally (iii) drawing motivation from AdS/CFT
we wish to explore systematically the freedom we have in choosing the boundary
conditions while maintaining (i) and (ii). The perspective our study brings is
that it illustrates the fact that there is not a unique boundary condition, but
rather a class of boundary conditions that all lead to a well defined notion of
asymptotic flatness. Various generalizations and variations on the study
presented below are possible. This study is a continuation of \cite{MM, MMV,
MMMV, CDV, V} and is largely motivated by comments in 
\cite{MM, Ma, MV}. For an alternative
point of view on some of these ideas see \cite{CD}. Other studies of
supertranslations include a series of papers by Barnich et
al \cite{Barnich:2010eb} at null infinity.

\setcounter{equation}{0}
\section{Asymptotic Flatness, Actions, Supertranslations}
\label{sec:review}
In this section we first provide relevant definitions and a brief review of previous work. We then spell out our boundary conditions. 

\subsection{Asymptotic Flatness and the Mann-Marolf Action}
As in previous work \cite{MM, MMV, MMMV, CDV, V}, we introduce our notion of
asymptotic flatness based on the work of Beig and Schmidt \cite{BS, B}. The
key advantage of using Beig-Schmidt expansion is that all results can be readily
translated to the geometrical language of Ashtekar and Hansen
\cite{Ashtekar:1978zz, AshRev} or that of Ashtekar and Romano \cite{Ashtekar:1991vb}.

Beig-Schmidt expansion for asymptotically flat spacetimes near spatial infinity
takes the form
\be
\label{metric1}
ds^2 = \left( 1 + \frac{\s}{\rho} \right)^2 d \rho^2 + \rho^2\left( h^{0}_{ab} +
\frac{h^{1}_{ab}}{\rho} + \frac{h^{2}_{ab}}{\rho^2} + {\mathcal O}(\rho^{-3})
\right) dx^a dx^b,
\ee
where $h^{0}_{ab}dx^ dx^b$ is the metric on the unit three-dimensional de Sitter
space $dS_3$, or equivalently on the unit three-dimensional hyperboloid
\cite{BS},
\be
h^{0}_{ab}dx^a dx^b = -d\tau^2 + \cosh^2\tau (d\theta^2 + \sin^2\theta d\phi^2).\label{h0}
\ee
We use  $\cD_a$ to denote the unique torsion-free covariant derivative
compatible with the metric $h^0_{ab}$ on the unit hyperboloid. The radial
 coordinate $\rho$ is associated to some asymptotically Minkowski coordinates
 $x^\mu$  via $\rho^2 = \eta_{\mu \nu} x^\mu x^\nu$. The fields $\sigma$,
$h^{1}_{ab},h^{2}_{ab}$, etc.~are assumed to be smooth functions on the unit
hyperboloid. We use $h_{ab}$ to denote the complete
induced metric on a constant $\rho$ slice (for some large $\rho$) and use $D$
to denote the unique torsion-free covariant derivative compatible with $h_{ab}$.
Further boundary conditions will be specified below.

Next, we recall the action principle of \cite{MM}. There it was shown that
a good variational principle for asymptotically flat configurations defined by
the expansion \eqref{metric1} is given by the action
\begin{equation}
\label{action} S = \frac{1}{16\pi G} \int_{\cal M} d^4x \sqrt{-g}\,R +
\frac{1}{8\pi G} \int_{\partial {\cal M}} d^3x\sqrt{-h} \,(K - \hat K).
\end{equation}
The counterterm in the action \eqref{action} is $\hat K :=
h^{ab} \hat K_{ab}$. $\hat K_{ab}$ is defined implicitly via a Gauss-Codacci
like equation
\begin{equation}
\label{defhatK}
{\cal R}_{ab} = \hat K_{ab} \hat K - \hat K_a{}^{c} \hat K_{cb}.
\end{equation}
Here ${\cal R}_{ab}$ is the Ricci tensor of the boundary metric $h_{ab}$. For details we refer the
reader to \cite{MM, MMV, MMMV, V}.

\subsection{Supertranslations}
In the introduction section we already mentioned certain basic facts about
supertranslations. The best way to further understand the precise nature of
supertranslations is to work out the set of diffeomorphisms  that preserve
the form of appropriately defined asymptotically flat metrics. Since results
obtained in the Beig-Schmidt coordinates \eqref{metric1} can be readily
translated to the geometrical languages, it is most natural to work out these
diffeomorphisms in this gauge. Supertranslations in the Beig-Schmidt gauge are
interpreted as different conformal completions in the SPI framework
\cite{Ashtekar:1978zz} or as different hyperboloid completions in the Ashtekar-Romano
framework \cite{Ashtekar:1991vb}.  The problem of finding these
diffeomorphisms has been analyzed by several authors \cite{BS, Ashtekar:1978zz, 
Ashtekar:1991vb, B}; for reviews see \cite{CDV, AshNew}. The
upshot of this analysis is that in an asymptotically Cartesian coordinate
system with $\rho^2 = \eta_{\mu \nu} x^{\mu} x^{\nu}$, all diffeomorphisms of
the form
\be
\bar{x}^\mu = L_{\nu}^{\mu}x^\nu + T^\mu + S^\mu(x^a)  + o(\rho^0) \label{diffeos}
\ee
preserve the form of the metric \eqref{metric1}. The transformations generated
by the constants $L_{\nu}^{\mu}$ and $T^\mu$ constitute the Poincar\'e group.
The transformations generated by angle dependent translations $S^\mu(x^a)$ are
the so-called supertranslations. In fact, they are all spi-supertranslations.
In the Beig-Schmidt expansion, the asymptotic spi-supertranslation Killing
vector $\xi^\mu_{\omega}$ is related to an arbitrary function $\omega$ on the
hyperboloid as
\be
\xi^{\rho}_\o = \omega(x) + \mathcal{O}(\rho^{-1}), \qquad
\xi^{a}_\o = \frac{1}{\rho} \o^a(x) + \mathcal{O}(\rho^{-2}),
\ee
where $\o^a = \cD^a \omega$ and where $x$ denotes collectively the coordinates
on the hyperboloid. As emphasized in the introduction, we  need an additional structure at
spatial infinity that can reduce the SPI group to the Poincar\'e group, as
otherwise the notion of angular momentum  is not the familiar one. Furthermore, since
supertranslations depend arbitrarily on the angular coordinates, in particular
on the time coordinate $\tau$, even if one attempts to define conserved
charge for them, the associated charges will in general be not conserved. Not
surprisingly, a large body of work on asymptotic flatness at spatial  infinity
 has taken the point of view to strengthen the boundary conditions. With the
 strengthened boundary conditions the freedom of performing
 supertranslations is eliminated. This is achieved in \cite{ABR,
 Ashtekar:1978zz, Ashtekar:1991vb, B,  AshNew} by demanding the leading order
 asymptotic Weyl curvature to be purely electric . We will continue to demand this condition.
There is still some freedom left and this is what we would like to draw the
attention of the reader to.

Let us look at the next to leading order asymptotic equations
of motion. It turns to be convenient to work with the variable
\be
k_{ab} = h^{1}_{ab} +2 \sigma h^{0}_{ab}\label{def:kab}.
\ee
Equations of motion at first order now take the form \cite{BS} 
\bea 
\square \sigma + 3 \sigma &=& 0, \label{EOMs} \\ 
\cD^b k_{ab} - \cD_a k &=& 0, \label{kabEOM1}\\ 
 (\square -3)k_{ab}
+k h^{0}_{ab} -\cD_a \cD_b k &=& 0 \label{kabEOM2},
\eea
where $\square = \cD^a \cD_a$ and $k = k_{a}{}^a$. 
The important point to note here is
that equations for the fields $\s$ and $k_{ab}$ are decoupled.
 By introducing the leading
order electric and magnetic parts of the Weyl tensor,
\be
E^{1}_{ab} = -\cD_a \cD_b \sigma - h^{0}_{ab} \sigma \, , \qquad \qquad
B^{1}_{ab} = \frac{1}{2} \: \curl k_{ab} = \frac{1}{2}\epsilon_a^{\;\; c d} \cD_c k_{d b} \,
\label{dBk},
\ee
one can rewrite
these equations in a more enlightening form. See, for example, reference
\cite{V} for a detailed discussion on this.  Our boundary conditions require
\be
\fbox{$\displaystyle
B^{1}_{ab} = 0.$}
\end{equation}
This implies that $k_{ab}$ must be of the form
\be
k_{ab} =2 \cD_a \cD_b \omega + 2h^{0}_{ab} \omega, \label{kabmain}
\ee
for some arbitrary $\omega$. 
This is because
the combination $\cD_a \cD_b \omega + h^{0}_{ab} \omega$ has vanishing curl,
and hence it does not contribute to the magnetic part of the Weyl tensor. For
the form \eqref{kabmain} of $k_{ab}$ equations of motion \eqref{kabEOM1} and
\eqref{kabEOM2} are also automatically satisfied. Now recall that this freedom
in the choice of $k_{ab}$ also exactly correspond to performing
supertranslations in the space of Beig-Schmidt configurations \eqref{metric1}.  By choosing a particular
representative for the inverse of the curl operator, i.e.,  a particular $\o$ in \eqref{kabmain},  
the freedom of performing supertranslations is eliminated.
Once such a choice is made, the function $\omega$ is fixed. It is best to regard it as a fixed background structure. This background structure is
precisely what we mean by the phrase `supertranslation frame.'  The most studied choice corresponds to taking $\omega = 0$ \cite{ ABR,
MMV, MMMV, Ashtekar:1978zz,  Ashtekar:1991vb, B}. We show in the rest of this
 section that other choices of $\omega$ are equally allowed. In this paper we
 wish to explore precisely the physics of making such a choice. To set the stage
 for this discussion, we first need to look at the boundary conditions of
 Ashtekar-Bombelli-Reula \cite{ABR}. These boundary
 conditions were in turn motivated by \cite{Ashtekar:1978zz}. We will comment on
 the motivation of references \cite{ABR, Ashtekar:1978zz} for choosing $\o =0$
 in section \ref{sec:conclusions}.

\subsubsection{Ashtekar-Bombelli-Reula Boundary Conditions}
For gravitational theories it is a well known fact that the boundary conditions
play a crucial role in the description of the phase space. For such
considerations it is often convenient to work in the covariant phase space
formalism  \cite{ABR, LW, W, IW, WZ}. The key quantity to consider in this
approach is the symplectic current vector $w^\mu$. The symplectic current vector
depends on the background metric $g$ and on perturbations around the background
metric $\delta_1 g$ and $\delta_2 g$.  It is skew symmetric
in the pair ($\delta_1 g$, $\delta_2 g$), and for the case of general relativity
it takes the form
\be
w^\mu = P^{\mu \nu \a \b \g \d} (\delta_2 g_{\nu \a} \nabla_\beta \delta_1 g_{\g
\d}- \delta_1 g_{\nu \a} \nabla_\beta \delta_2 g_{\g
\d}),
\ee
where
\be
P^{\mu \nu \a \b \g \d} = 
g^{\mu \g} g^{\d \nu} g^{\a \b} 
- \frac{1}{2}g^{\mu \b} g^{\nu \g} g^{\d \a}  
- \frac{1}{2}g^{\mu \nu} g^{\a \b} g^{\g \d}  
- \frac{1}{2}g^{\nu \a} g^{\mu \g} g^{\d \b}  
+ \frac{1}{2}g^{\nu \a} g^{\mu \b} g^{\g \d}. 
\ee
Using the Ansatz
\be
ds^2 = g_{\mu \nu} dx^\mu dx^\nu = N^2 d \rho^2 + h_{ab} dx^a dx^b,
\label{ansatz1}
\ee
and
\be
\delta ds^2 = \delta g_{\mu \nu} dx^\mu dx^\nu = 2 N \delta N d \rho^2 +
\delta h_{ab} dx^a dx^b,
\label{ansatz2}
\ee
we obtain the 3+1 split of the symplectic current vector. The radial component
reads
\bea
w^\rho &=& \frac{1}{4N^3} h^{ab}h^{cd} \Bigg{\{} \Big{[} N h^{ef}  \delta_1
h_{ab} \delta_2 h_{ce} \partial_\rho h_{df} 
+ 2 \delta_2 h_{ac}(\delta_1 N
\partial_\rho h_{bd} - N \partial_\rho \delta_1 h_{bd}) \nn \\ 
&&- 2 \delta_2 h_{ab}(\delta_1 N \partial_\rho h_{cd} - N \partial_\rho
\delta_1 h_{cd}) \Big{]} - (1 \leftrightarrow 2) \Bigg{\}}, \label{wrho}
\eea
whereas the angular components read
\bea
w^f  &=& \frac{1}{2N}  h^{fa} h^{bc} \Bigg{\{} 
  2 \delta_2 N D_a \delta_1 h_{bc} +
  2 \delta_2 h_{bc} D_a \delta_1 N 
- 2 \delta_2 N D_c \delta_1 h_{ab}
- 2 \delta_2 h_{ab} D_c \delta_1 N \nn \\ & & 
+ h^{de} \Big{[} 
 N \delta_2 h_{bc} D_a \delta_1 h_{de}
- N \delta_2 h_{bd} D_a \delta_1 h_{ce} 
- \delta_1 h_{ab} \delta_2 h_{de} D_c N
- N \delta_2 h_{ab} D_c \delta_1 h_{de} \nn \\ & & 
+ 2 N \delta_2 h_{bd} D_e \delta_1 h_{ac}
- N \delta_2 h_{bc} D_e \delta_1 h_{ad}
\Big{]}  - (1 \leftrightarrow 2) \Bigg{\}}.
\label{wf}
\eea
For the Ansatz \eqref{ansatz1} and \eqref{ansatz2},
equations \eqref{wrho} and \eqref{wf} are general expressions for the
radial and the tangential components of the symplectic current $w^\mu$. In
arriving at these expressions no reference to any boundary conditions has been
made.
Although these equations look somewhat clumsy, from the computational point of
view these are the easiest expressions to work with.

The integral of the Hodge dual of the symplectic current vector over a Cauchy
slice $\Sigma$ defines the symplectic structure. One must choose boundary conditions
to ensure that the symplectic structure is finite and conserved. When
$\delta_1 g $ and $\delta_2 g$ satisfy linearized equations of motion, it
follows from a standard argument that $\nabla_\mu w^{\mu} =0$, where
$\nabla_\mu$ is the covariant derivative compatible with the bulk metric $g$.
Therefore, the two requirements---finiteness and conservation of the symplectic
structure---reduce to respectively
\be
\frac{1}{16\pi G}\int_\Sigma \star_4 w^\mu < \infty \qquad \mbox{and}  \qquad
\frac{1}{16\pi G}\int_{\Sigma_{12}} \star_4 w^\mu = 0.
\ee
Here, $\star_4$ denotes the four-dimensional Hodge star, and surface
$\Sigma_{12}$ is defined as follows. Let $\Sigma_1$ and $\Sigma_2$ be two Cauchy
surfaces ending at spatial infinity. These surfaces enclose a spacetime volume
bounded by $\Sigma_1$ and $\Sigma_2$ and a portion of the boundary.
$\Sigma_{12}$ denotes that portion of the boundary. With our notion of
asymptotic flatness these requirements are translated into, respectively,
\be
\frac{1}{16\pi G}\lim_{\rho \to \infty}\int_\Sigma \sqrt{-g} w^{\tau}  d \rho d
\theta d \phi < \infty \qquad \mbox{and}  \qquad \frac{1}{16\pi G} \lim_{\rho
\to \infty} \int_{\Sigma_{12}} \sqrt{-g} w^\rho d\tau d\theta d\phi = 0,
\ee
where we have taken Cauchy surfaces $\Sigma_{1,2}$ to asymptote to constant
$\tau$ surfaces in the hyperboloid. Ashtekar, Bombelli, and Reula showed that
with the boundary conditions
\bea
\quad h^1_{ab} &=&- 2 \s h^0_{ab},\quad 
\label{ABR1}\\
\quad \delta h^1_{ab} &=&- 2 \delta \s h^0_{ab},\quad  
\label{ABR2}
\\
\quad \sigma(\tau, \theta, \phi)&=& \s(-\tau, \pi - \theta, \phi + \pi),\quad
\label{ABR3}
\label{sigmasymm}
\eea
both the above requirements are satisfied. In particular, for the integral
\be
\frac{1}{16\pi G}\lim_{\rho \to \infty}\int_\Sigma \sqrt{-g} w^{\tau}  d \rho d
\theta d \phi 
= \frac{1}{4\pi G}\lim_{\rho \to \infty}\int_\Sigma\sqrt{-h^0}
\frac{1}{\rho} \left(\delta_1  \s \cD^\tau \delta_2 \s - \delta_2 \s \cD^\tau
\delta_1 \s \right)d\rho d \theta d
\phi,
\ee
one find that the potentially divergent term on the right hand side vanishes
upon using boundary condition \eqref{sigmasymm}.  In the boundary conditions
\eqref{ABR1}--\eqref{ABR3} the choice $\omega = 0$ has been
made. Since a particular choice has been made,  supertranslations do not
act on the phase space.

\subsubsection{Supertranslated Boundary Conditions}
The boundary conditions we work with in this paper are as follows
\begin{empheq}[box=\widefbox]{align}
\quad h^1_{ab} &=- 2 \s h^0_{ab} + 2 \cD_a \cD_b \omega + 2 h^0_{ab} \o,\quad 
\label{sABR1} \\ 
\quad \delta h^1_{ab} &=- 2 \delta \s h^0_{ab},\quad  \label{sABR2} \\
\quad \sigma(\tau, \theta, \phi)&= \s(-\tau, \pi - \theta, \phi + \pi).\quad
\label{sABR3}
\end{empheq}
In the boundary conditions
\eqref{sABR1}--\eqref{sABR3} the choice $\omega \neq 0$ has been
made. Once again since a particular choice for $\omega$ has been made, 
 supertranslations \emph{do not} act on the phase space.
Nothing changes in the calculation of the symplectic structure when working with
these boundary conditions. The symplectic structure is still finite and
conserved as is the case with the ABR boundary conditions.  It is best to regard
$\o$ as the fixed background structure. We refer to boundary conditions \eqref{sABR1}--\eqref{sABR3} as
the supertranslated ABR boundary conditions. 

At this point we wish to point out that such a generalization
should be possible was already speculated in the work of Mann and Marolf
\cite{MM}, though the precise boundary conditions 
\eqref{sABR1}--\eqref{sABR3}
were not stated.

\setcounter{equation}{0}
\section{Asymptotic Expansions}
\label{sec:expansion}
Having specified our boundary conditions we now wish to study the consequences
on the asymptotic equations of motion and on the construction of the boundary
stress tensor at the next-to-next-to leading order. It is necessary to work at
this order to get a handle over the construction of Lorentz charges.  We will concentrate mostly on the
physics, and will not go into much calculational details. Since we carry out
asymptotic expansions at second order for arbitrary $\o$,  the manipulations
involved are in fact quite intricate and tedious.
\subsection{Second Order Equations of Motion}
The radial 3+1 split of the bulk Einstein equations give the following equations
for the Ansatz \eqref{ansatz1} \cite{B}
\bea
h^{ab} \partial_\rho K_{ab} - N K_{ab}K^{ab} + D^2 N &=& 0, \\
D_b K^b{}_{a} - D_a K &=& 0, \\ 
\cR_{ab}- N^{-1} \partial_\rho K_{ab} - N^{-1} D_a D_b N - K K_{ab} +
2 K_{a}{}^{c} K_{cb} &=& 0.
\eea
Here $K_{ab}$ denotes the extrinsic curvature of the constant $\rho$
hypersurface. We carry out the expansion of these equations systematically in
appendix \ref{app:EOM}. The final outcome of this analysis is the second
order equations of motion. These equations take the form
\bea
h^2 &=& 12 \s^2 + \s_a \s^a + 3 \omega^2 + 2 \omega \Box \omega +
\omega_{ab}\omega^{ab} - 9 \omega \s - \s \Box \omega + \s_a \omega^a + \s_a
\Box \omega^a \nn \\
& & + 2 \s_{ab}\omega^{ab},\label{traceh2}\\
\cD^b h^2_{ab} &=& 16 \s \s_a + 2 \s_{ab}\s^b + 2 \omega \omega_a + 2 \omega
\Box \omega_a + 2 \omega^b \omega_{ab} + \omega_{ab}\Box \omega^b +
\omega_{abc}\omega^{bc} \nn \\
& & - \s \omega_a - 3 \omega \s_a + \s_a \Box \omega - \s \Box \omega_a + 3
\s_{ab} \omega^b - \o_{ab} \s^b + \s_{ab}\Box \o^b \nn + \s^b \Box \o_{ab} \\ 
&& 
+ 2 \s_{abc}\o^{bc} + 2 \o_{abc} \s^{bc}, \label{divh2}\\
(\Box - 2 )h^2_{ab} &=& 	6(\s_c \s^c - 3 \s^2)h^0_{ab} + 8 \s_a \s_b + 14 \s
\s_{ab} + 2 \s_{ac}\s^{c}{}_{b} + 2 \s_{abc}\s^c\nn \\
& & + 2  (\omega \Box \o - \o^2 + \o_c \o^c )h^0_{ab} 
 - 4 \o \o_{ab} + 2 \o_{ab}\Box \o + 2 \o \Box \o_{ab}
 \nn \\ &&
  + 4 \o_{abc}\o^c
-2 \o_{cb}\o^{c}{}_{a} + 2 \o_{a}{}^{cd} \o_{bcd} + 2 \o_{c(a} \Box \o_{b)}{}^c
\nn \\
& & + (14 \o \s - 4 \s \Box \o - 4 \s_c \o^c + 2 \s^c \Box \o_c + 4
\s_{cd}\o^{cd})h^0_{ab} + 17 \s \o_{ab} - \o \s_{ab} \nn \\
& & 
- \s_{ab}\Box \o - \s \Box \o_{ab} + 5 \s_{abc}\o^c - 5 \s^c \o_{abc} + \s_{abc}
\Box \o^c + \s^c \Box \o_{abc} \nn \\
& &  + 2 \s_{abcd}\o^{cd} + 2 \o_{abcd}\s^{cd} + 2 \s_{c(a} \o_{b)}{}^c  + 2
\s_{c(a} \Box \o_{b)}{}^c + 4 \s_{(a}{}^{cd} \o_{b)cd}. 
\label{boxh2}
\eea
In writing these equations we use the following compact 
notation,
\bea
\o_{abcd} &=& \cD_d \cD_c \cD_b \cD_a \o, \\
\Box \o_{abc} &=& (\cD^e \cD_e) \o_{abc} = \cD^e \cD_e \cD_c \cD_b \cD_a \o,
\eea
etc.~and similarly for $\s$.
In the special case when $(\Box + 3) \omega = 0$ these equations can
also be extracted from \cite{CD}.

\subsection{Integrability Conditions}
The second order equations of motion \eqref{traceh2}--\eqref{boxh2} are in fact
quite complicated. It might seem difficult to rewrite these equations in a form that can be used to perform
an integrability analysis following Beig \cite{B}. Remarkably enough, this is
not the case. These equations have a somewhat magical structure: all the $(\s,\omega)$
terms and all $(\omega,\omega)$ terms on the right hand side of equation
\eqref{boxh2} can be repackaged as $(\Box -2)$ acting of the following tensor 
\bea
\chi_{ab}& = & - \s \o_{ab} - \o \s_{ab} - 4 h^0_{ab} \s \o
+ 2 h^0_{ab} \s_c \o^c + \s_{abc}\o^c + \o_{abc}\s^c + 2
\s_{c(a}\o_{b)}{}^{c} \nn \\
& & 
+ 2 \o \o_{ab} + \omega_a{}^c
\omega_{bc} + h^0_{ab} \o^2 \label{chiab}
\eea
As a result \eqref{boxh2} can be written as
\bea
(\Box -2)(h^2_{ab} - \chi_{ab}) = 6(\s_c \s^c - 3 \s^2)h^0_{ab} + 8 \s_a \s_b + 14 \s
\s_{ab} + 2 \s_{ac}\s^{c}{}_{b} + 2 \s_{abc}\s^c\nn.
\eea
The usefulness of the tensor
$\chi_{ab}$ goes well beyond that. Equation \eqref{traceh2} can be rewritten
as
\be
h^2 - \chi = 12 \s^2 + \s_a \s^a,
\ee
and similarly the divergence equation \eqref{divh2} is rewritten as 
\be
\cD^b(h^2_{ab} - \chi_{ab})= 16 \s \s_a + 2 \s_{ab}\s^b.
\ee
Written in this form the second order equations are much more manageable. Now,
following the discussion in \cite{CDV} we define a tensor $V_{ab}$ as
\bea
V_{ab}= -h^2_{ab} + \chi_{ab} + 6 \s^2 h^0_{ab} + 2 \s_{ab} \s - 2 \s_a \s_b +
\s^c \s_c h^0_{ab}.
\eea
In terms of $V_{ab}$ the equations of motion take the form
\begin{empheq}[box=\widefbox]{align}
 V_{a}^{a} &=0,\\ 
 \cD^aV_{ab} &=0, \\
(\Box-2)V_{ab}&= \curl [ 4 \epsilon_{cd(a} \s^c \s^{d}_{b)}],
\end{empheq}
where as in \eqref{dBk} $\curl$ of a tensor $T_{ab}$ is defined as 
\be
\curl T_{ab} = \epsilon_{a}{}^{cd} \cD_c T_{db}. 
\ee
For further properties of the $\curl$ operator and of the tensor
structure $\epsilon_{cd(a} \s^c \s^{d}_{b)}$ we refer the reader to \cite{CDV}. 

Since $V_{ab}$ is symmetric, traceless, and divergence free,  discussion of
the integrability conditions of \cite{CDV} applies as is. We find
 that despite the fact that we have an arbitrary function $\omega$ in our
asymptotic expansion the integrability conditions do not change.
The integrability conditions require Lorentz charges constructed using 
\be 
\curl [ 4
\epsilon_{cd(a} \s^c \s^{d}_{b)}]
\ee 
to be zero. In this paper, we have chosen
 the mass aspect $\sigma$ to be a symmetric function on the hyperboloid. As a
 result, the integrability conditions are automatically satisfied
 \cite{MMMV, CDV}. The outcome of this is that the tensor $V_{ab}$ can be
 readily used to construct well defined and conversed Lorentz charges\footnote{One comment regarding tensor $\chi_{ab}$ is in order here: the form of $\chi_{ab}$ \eqref{chiab} can also be worked out by calculating the
non-linear action of supertranslation $\o$ on $h^2_{ab}$ starting with the ABR
boundary conditions, in particular using equation \eqref{ABR1}. This calculation in a
somewhat different context was first performed in an unpublished work in
collaboration with  Geoffrey Compere and Francois Dehouck. For the special case
when $\cD^a k_{ab} = k_{a}^{a} = 0$, i.e., $(\Box + 3)\o = 0$, such an
expression can also be extracted from \cite{CD}. I thank Geoffrey Compere and
Francois Dehouck for their permission to use material from this joint
unpublished work.}.

\subsection{Expansion of Counterterm}
\label{sec:expansionCounterterm}
Having analysed the second order equations of motion and the integrability
conditions, we now turn to the expansion of the Mann-Marolf counterterm.  Recall that the counterterm $\hat K$
is defined implicitly via the Gauss-Codacci like equation \eqref{defhatK}. It
is convenient to introduce $\hat p_{ab} = \frac{1}{\rho} \hat K_{ab}$.
Expanding
$\hat p_{ab}$ as
\begin{equation}
\hat p_{ab} = h^0_{ab} + \frac{1}{\rho} \hat p^1_{ab} +
\frac{1}{\rho^2} \hat p^2_{ab} + {\cal O} \left(
\frac{1}{\rho^3}\right), \label{expansionp}
\end{equation}
we can invert the relation \eqref{defhatK} and  express $\hat p^1_{ab}, \hat
p^2_{ab}$ in terms of the expansion of the Ricci tensor on the hyperboloid.
This computation was first done in \cite{MMV} for the ABR boundary conditions. We refer the reader to the
appendix B of \cite{MMV} for details.
By a direct calculation we find upon using equations of motion obtained above
\begin{eqnarray}
\hat p^1_{ab} = \s_{ab} - \s h^0_{ab} + \o_{ab} + h^0_{ab} \o.
\end{eqnarray}
A similar calculation for $\hat p^2_{ab}$ gives
\bea
\hat p^{2}_{ab} &=& h^{2}_{ab} - \left( \frac{5}{4}\s^2 + \s_c \s^c + 
\frac{1}{4} \s^{cd}\s_{cd}\right) h^0_{ab}  +
2 \s_a \s_b + \s \s_{ab} + \s_a{}{}^c \s_{cb}  
- h^0_{ab} \o^2  \nn \\ && - 2 \o \o_{ab}  - \o_{a}{}^{c} \o_{cb} 
+ \left(3  \s \o 
+ \frac{3}{2} \s \Box \o - \s^c \o_c + \frac{1}{2}
\s_{cd}\o^{cd} \right) h^0_{ab} 
+ \o \s_{ab} 
 \nn \\ &&
+ \s_{ab} \Box \o 
- \o_{abc}\s^c  - 2 \s_{c(a}\o_{b)}{}^{c}.
\label{phat2:1}
\eea
The traces of $\hat p^{1}_{ab}$ and $\hat p^{2}_{ab}$ simplify to
\bea
\hat p^1 &:=& h^{0 ab}\hat p^{1}_{ab} = - 6 \s + \Box \o + 3 \o,\\
\hat p^2 &:=& h^{0 ab}\hat p^{2}_{ab} =  \frac{21}{4}\s^2 + \frac{1}{4} \s_{cd}
\s^{cd} - 3 \o \s + \frac{1}{2} \s \Box \omega + \frac{3}{2} \s_{cd}
\o^{cd}.
\label{tracephat2}
\eea
These equations are important for the considerations of the next section.

\setcounter{equation}{0}
\section{Supertranslations and Boundary Stress Tensor}
\label{sec:BST}
In this section we study the on-shell value of the action and its first
variations. We also compute the next to leading order expression for the
boundary stress tensor. We follow the corresponding discussion in \cite{MM,
MMMV, V}. The new element in the following discussion is our boundary conditions
\eqref{sABR1}--\eqref{sABR3}.

\subsection{First Variations}
Let us consider the first variations of the Mann-Marolf action over
configurations satisfying our boundary conditions \eqref{sABR1}--\eqref{sABR3}
and evaluate it on-shell.
This set-up was already considered in \cite{MM} so we shall be brief. The first
variation of the Mann-Marolf action is
\cite{MM, MMMV, V}
\be
(16 \pi G) \delta S_{\rom{total}} = \int_{\partial \cM} \sqrt{-h} d^3 x(\pi^{ab} - \hat \pi^{ab} + \Delta^{ab}) \delta h_{ab},
\ee
where $\pi^{ab} =  K h^{ab}- K^{ab}$, $\hat \pi^{ab} = \hat K
h^{ab}- \hat K^{ab}$ and $\Delta_{ab}$ is
\be
\Delta^{ab} = \hat K^{ab} - 2 \tilde L^{cd} (\hat K_{cd} \hat K^{ab} - \hat
K^{a}_{c} \hat K_{d}^{b}) + D^2 \tilde L^{ab} + h^{ab}D_c D_d \tilde L^{cd} - 2
D_{d}D^{(a}\tilde L^{b)d},
\label{Delta}
\ee
with $L_{ab}{}^{cd}$ and $\tilde L^{ab}$ given by \cite{MM, MMMV, Ross}
\be
L_{ab}{}^{cd} = h^{cd}\hat K_{ab} + \delta_{(a}^c\delta_{b)}^d \hat K -
\delta_{(a}^c\hat K_{b)}^d- \delta_{(a}^d\hat K_{b)}^c, \qquad \quad \tilde
L^{ab} := h^{cd}(L^{-1})_{cd}{}^{ab}.
\ee
Using asymptotic expansions of the previous section it follows that
\be
(\pi^{ab} - \hat \pi^{ab} + \Delta^{ab}) = \frac{1}{\rho^{4}}\left(\s^{ab} + \s
h^{0ab}\right) + \mathcal{O}\left(\frac{1}{\rho^{5}}\right).
\label{inter}
\ee
Now, using our boundary condition \eqref{sABR2} we see that  in
the $\rho \to \infty $ limit
\be
(16 \pi G) \delta S_{\rom{total}} = \int_{dS_3} \sqrt{-h^{0}} d^3 x\left(
\s^{ab} + h^{0\:ab}\s\right)\left(- 2 \delta \s h^0_{ab}\right).
\ee
The equation of motion for $\sigma$ now immediately tells us that the
first variation of the action vanishes identically
\be
\delta S_{\rom{total}} = 0.
\ee
Thus, action \eqref{action} provides a good
variational principle for our notion of asymptotic flatness\footnote{Alternatively, using
$\delta h_{ab} = \rho \delta h^{1}_{ab} + \ldots = - 2 \rho \delta \s
h^{0}_{ab} + 2 \rho \cD_a \cD_b \delta \o + 2 \rho \delta \o h^0_{ab}+ \ldots$
and $\sqrt{-h} = \rho^{3}\sqrt{-h^{0}} + \ldots $ it follows that in the $\rho
\to \infty $ limit
\be
(16 \pi G) \delta S_{\rom{total}} = \int_{dS_3} \sqrt{-h^{0}} d^3 x\left(
\s^{ab} + h^{0\:ab}\s\right)\left(- 2 \delta \s h^0_{ab} + 2  \cD_a \cD_b
\delta \o +  2 \delta \o h^0_{ab}\right).
\ee
Using the equation of motion
for $\sigma$, this equation further simplifies to
\be
(16 \pi G) \delta S_{\rom{total}} = \int_{dS_3} \sqrt{-h^{0}} d^3 x\left(
\s^{ab} + h^{0\:ab}\s\right)\left( 2  \cD_a \cD_b
\delta \o\right).
\ee
Performing integration by parts and using equation of motion for $\s$ one more
time, we see that the first variation of the action vanishes identically
\be
\delta S_{\rom{total}} = 0.
\ee
In particular, supertranslations need not be fixed! Asymptotically
flat metrics related to each other via arbitrary supertranslations can be
consistently considered in the Mann-Marolf variational principle. See also
\cite{CD}. However, this is not the boundary conditions we use in this paper
for reasons emphasized in the introduction section.
\label{footnote:supertranslation}}.

\subsection{On-shell Action}
\label{onshell}
We now calculate the on-shell value of the action. Given our results above, 
this calculation is rather straightforward. Since our
spacetimes are Ricci flat the bulk term in \eqref{action} vanishes on-shell.
Therefore,
\be
 S_{\rom{on-shell}} = \frac{1}{8 \pi G} \int_{\partial \cM}
d^3x\sqrt{-h} (K - \hat K).
\ee
Now, using expansions of $K_{ab}$  and $\hat K_{ab}$ obtained above (section
\ref{sec:expansionCounterterm} and Appendix \ref{app:EOM} respectively)  we
have
\be
 S_{\rom{on-shell}} =  \frac{1}{32
\pi G} \int_{dS_3}d^3x\sqrt{-h^{0}}\left[3 \s^2 - \s_{ab}\s^{ab} + 2 \s \Box \o + 2
\s_{ab}\o^{ab} \right].
\label{onshell}
\ee
All divergent terms have cancelled. The on-shell is finite. Doing integrations
by parts and using equations of motion for $\s$, we observe that the on-shell
action vanishes
\be
S_{\rom{on-shell}} = 0.
\ee
In particular, the on-shell value \emph{does not} depend on the supertranslation
frame $\omega$.
An interpretation of this result is as follows \cite{V}. We showed above that
$\delta S = 0$ on all variations satisfying our boundary conditions. It follows
that $S_{\rom{on-shell}}$ must be constant as we move along any smooth
path in our phase space. Furthermore, we expect all
configurations to be smoothly connected to Minkowski space. For Minkowski space
$S_{\rom{on-shell}}$ is identically zero. Therefore,  it follows that
$S_{\rom{on-shell}}$ is identically zero on any asymptotically flat solution satisfying our boundary condition. For more
comments on this point see \cite{V} and also footnote
\ref{footnote:supertranslation}.

\subsection{Boundary Stress Tensor}
From the first variation of the action, the boundary stress tensor can also be
computed. It admits an expansion in the inverse powers of $\rho$. The leading
order and the next to leading order terms in the expansion are relevant for the
construction of translations and Lorentz charges respectively \cite{MM, MMV}.
After a long and tedious computation we find these expressions to be
\bea
T_{ab} = -\frac{1}{8 \pi G} \left( T^{1}_{ab} + \frac{1}{\rho}T^{2}_{ab}
+ \ldots \right) \label{Texpansion}
\eea
where
\bea
T^1_{ab} = \s_{ab} + h^0_{ab} \s \label{T1}
\eea
and
\bea
T^2_{ab} &=& h^2_{ab} + 2 \s_a \s_b + \frac{49}{4}\s \s_{ab} + 4 \s_{abc}\s^c +
7 \s_{a}{}^{c}\s_{bc} - \frac{3}{4} \s_{abcd}\s^{cd} +
\frac{9}{4}\s_{acd}\s_{b}{}^{cd} \nn \\
& & + \left[ 
\frac{35}{4}\s^2 + 3 \s_c \s^c - \frac{13}{4} \s_{cd}\s^{cd} -
\frac{3}{4}\s_{cde}\s^{cde} \right] h^0_{ab} + \frac{1}{2} \o_a \o_b - 2 \o
\o_{ab} + \o_{ab} \Box \o \nn \\
& & + \o_{(a}\Box \o_{b)} + \o_{abc}\o^{c} - 4 \o_{ac}\o_b{}^{c} +
\frac{1}{2}\Box \o_a \Box \o_b - \frac{1}{2} \Box \o \Box \o_{ab} - 2
\o_{abc}\Box \o^{c} \nn \\
& &+ 2 \o_{c(a}\Box \o_{b)}{}^{c} - \o_{ab} \Box \Box \o -
\frac{1}{2}\o_{abcd}\o^{cd} + \frac{3}{2} \o_{acd}\o_{b}{}^{cd} +
\left[-\o^2 + \frac{1}{2} \o_c \o^c - 2 \o_c \Box \o^{c} 
\right. \nn \\
&& \left.
+ \frac{1}{2} \Box
\o^{c} \Box \o_{c}- \frac{1}{2}\o_{cd} \Box \o^{cd} + \frac{1}{2} \Box \o \Box
\Box \o - \frac{1}{2}\o_{cde}\o^{cde} \right] h^0_{ab} +  \s \o_{ab} + \o
\s_{ab} \nn \\
& & + \frac{7}{4} \s_{ab} \Box \o - \frac{9}{4} \s \Box \o_{ab} -
\frac{3}{2}\s_{abc}\o^{c} - \frac{11}{2} \o_{abc} \s^c - 2 \s_{c(a}
\o^{c}{}_{b)} + 3 \Box \o^c \s_{abc} - 3 \s^{c}{}_{(a} \Box \o_{b)c} \nn \\
& & + \frac{3}{2} \s_{ab} \Box \Box \o + \frac{3}{4} \s_{abcd}\o^{cd}+ 
\frac{3}{4} \o_{abcd}\s^{cd} - \frac{9}{2} \s_{(a}{}^{cd} \o_{b)cd} + \left[
4 \o \s + \frac{17}{4} \s \Box \o - \frac{11}{2} \s_c \o^{c} 
\right.
\nn \\
& & 
\left. 
+ \frac{9}{2} \s^c
\Box \o_c + \frac{9}{4} \s \Box \Box \o + \frac{13}{4} \s_{cd}\o^{cd} +
\frac{3}{4} \s_{cd}\Box \o^{cd} + \frac{3}{2} \s_{cde}\o^{cde}  \right]
h^0_{ab}.
\label{mainT}
\eea
Equation \eqref{mainT} is one of the main result of this paper. Certain
calculational details on how we obtained this expression can be found in
appendix \ref{app:BST}.

\setcounter{equation}{0}
\section{Properties of Boundary Stress Tensor}
\label{sec:BSTprop}
In this section we explore properties of our boundary stress tensor
\eqref{Texpansion}--\eqref{mainT}.
\subsection{Boundary Stress Tensor is Conserved a la Brown-York}
\label{sec:BSTprop1}
The above stress tensor can be shown to be conserved
\be
D^{b}T_{ab} = 0. \label{conserve} 
\ee
However, care must be exercised in interpreting this result. The
derivative $D^{a}$ in \eqref{conserve} is the torsion-free covariant derivative
compatible with the \emph{full metric on the hyperboloid $h_{ab}$.} When
expanded in powers of $\rho$ this equation reads at leading order
\be
\cD^{b}T^{1}_{ab} = 0,
\ee
and at the next to leading order
\be
\cD^{b}T^2_{ab} - \s \s_a - \s_{ab} \s^b - 2 \s \o_a 	- 2 \s \Box \o_a - 2
\s_{ab}\o^b - 2 \o_{ab}\s^b - \s_{ab} \Box \o^b - 2 \s_{abc}\o^{bc} -
\o_{abc}\s^{bc}=0. \label{conserve1}
\ee
An important question to ask at this point  is whether or not the
above expression can be written as a total derivative of a symmetric tensor $\tilde T_{ab}$. For
$\o$ independent terms this is indeed the case \cite{MMV}
\be
\tilde T_{ab} = T^2_{ab} - \s T^1_{ab}. \label{eq:Td}
\ee
When $\o$ dependent terms are included, with our preliminary investigations we
were unable to write \eqref{conserve1} as a total derivative of a symmetric
tensor. This is not necessarily an obstacle for the construction of conserved
charges. We already know from our study of the integrability conditions of the
second order equations of motion that a conserved tensor constructed using
$h^2_{ab}$---namely $V_{ab}$---exist and can be used to construct
conserved Lorentz charges.
We expect such a tensor to play an important role in the covariant phase space
construction of charges. Given the analysis of \cite{CD} and our considerations of the
covariant phase space above, it is fairly clear that such a construction goes
through without surprises. It can be interesting to fill in all details. We
will not pursue this direction here. On the other hand, construction of
conserved Lorentz charges using the boundary stress tensor approach is more
interesting and perhaps more difficult; we explore certain aspect of this in the
rest of the paper.

Reference \cite{MM} presented a general construction of boundary stress tensor charges starting with equation \eqref{conserve}.
There an expression for conserved charge for an asymptotic Killing vector 
$\xi_\rho^a$ is given in terms of the variation of the renormalized action
\be
Q[\xi] = - \Delta_{f, \xi} S_\rom{renorm} = - \lim_{\rho \to \infty}
\frac{1}{2} \int_{\partial \cM_{\rho}} \sqrt{-h} T^{ab} \Delta_{f, \xi_\rho}
h_{ab} d^3x,
\label{charge}
\ee
where 
\be
\Delta_{f,\xi} h_{ab} = (\pounds_{f\xi} g)_{ab} - f (\pounds_{\xi} g)_{ab},
\label{Deltafxi}
\ee
and where $f$ is smooth function that take the value $f=0$ at the past
boundary of $\partial \cM_{\rho}$ and the value $f=1$ at the future boundary. 
The right hand side of \eqref{Deltafxi} denotes quantities evaluated in
the bulk $\cM_{\rho}$ and then pulled back to the boundary $\partial
\cM_{\rho}$. Using general arguments it has been shown in \cite{HIM} that
this charge is also the generator of the asymptotic symmetry $\xi_\rho^a$.
Upon performing integrations by parts, equation \eqref{charge} can be converted
into an integral over a co-dimension two surface $C_\rho$---a cut in boundary
$\partial \cM_\rho$
\be
Q[\xi]  = \lim_{\rho \to \infty}
 \int_{\cC_{\rho}} \sqrt{-h_{C_\rho}} T_{ab} \xi_\rho^a n_\rho^{b} d^2x.
\label{charge2}
\ee

At this stage the above expression for conserved charges is somewhat formal. All
quantities that enter into this expression admit expansions in inverse powers of
$\rho$. For analysing Lorentz charges second order expansion of various
quantities is required, which makes the analysis quite intricate. 
Nevertheless, we expect that our boundary stress tensor can be used to construct
conserved charges. The precise details as to how this construction proceeds is
not investigated at this stage. We will return to this problem elsewhere in the future.

It is worthwhile to point out that for the case $\o = 0$ the
corresponding construction was carried out in \cite{MMV}, where divergence
free nature of tensor $\tilde T_{ab}$ \eqref{eq:Td} was also observed.
 Although it is fairly non-trivial to carry out explicit
construction of conserved charges 
for $\omega \neq 0$ in all detail, it is rather straightforward to study
transformation properties of Lorentz charges under translations from \eqref{charge2}.
We present this study in section \ref{sec:transform}. For now let us explore
some further properties of our stress tensor.

\subsection{Special Cases}
\label{sec:BSTprop2}
In this subsection we look at various special cases where our stress tensor
simplifies. In all cases it satisfies expected properties. This
study allows us to probe the structure of our stress tensor.
\subsubsection{$\omega = 0$}
When we choose $\omega = 0$ the boundary stress reduces to a previously computed
expression \cite{MMMV}
\bea
T_{ab} = -\frac{1}{8 \pi G} \left( T^{1}_{ab} + \frac{1}{\rho}T^{2}_{ab}
+ \ldots \right),
\eea
where
\bea
T^1_{ab} = \s_{ab} + h^0_{ab} \s,
\eea
and
\bea
T^2_{ab} &=& h^2_{ab} + 2 \s_a \s_b + \frac{49}{4}\s \s_{ab} + 4 \s_{abc}\s^c +
7 \s_{a}{}^{c}\s_{bc} - \frac{3}{4} \s_{abcd}\s^{cd} +
\frac{9}{4}\s_{acd}\s_{b}{}^{cd} \nn \\
& & + \left[ 
\frac{35}{4}\s^2 + 3 \s_c \s^c - \frac{13}{4} \s_{cd}\s^{cd} -
\frac{3}{4}\s_{cde}\s^{cde} \right] h^0_{ab}.
\label{simpleT}
\eea
Properties of this expression are already well studied in the
literature \cite{MMMV, CDV}.

\subsubsection{$\omega_{ab} + h^0_{ab}\omega = 0$}
When $\omega_{ab} + h^0_{ab}\omega = 0$, i.e., when $\omega$ is a translation,
$k_{ab}$ \eqref{kabmain} vanishes identically. In this case the asymptotic
metric expansion also reduces to the previously studied case of \cite{MMV, MMMV,
CDV}.
Therefore, we expect again the boundary stress tensor to reduce to \eqref{simpleT}. It can be
verified by a direct calculation that this is indeed the case. Remarkable cancellations happen when $\omega_{ab} +
h^0_{ab}\omega = 0$ is substituted in \eqref{mainT}. All $\omega$ dependent
terms reduce to zero, giving us \eqref{simpleT} as the final expression. This
provides a highly non-trivial test on our computations.

\subsubsection{$\sigma = 0$}
Another non-trivial case is when the mass aspect is
set to zero. In this case all conserved charges corresponding to translations
 vanish identically. Perhaps Minkowski space is the only solution with this
 property. In this section we wish to understand
 properties of the Lorentz charges when $\sigma =0$. When $\s$ is set to zero,
 $h^2_{ab}$ is solved from equations \eqref{traceh2}--\eqref{boxh2} to\footnote{with the
 most natural choice $V_{ab} =0$. A choice is necessary because the
 corresponding equations are hyperbolic.} read
 \be
 h^2_{ab} = 2 \o \o_{ab} + \o_{a}{}^{c}\o_{bc} + h^0_{ab} \o^2.
 \label{eqsubs}
 \ee
 Below we substitute this expression
 of $h^2_{ab}$ in the stress tensor. The resulting stress tensor is the
 stress tensor of Minkowski space in a general supertranslation gauge. The fact that the following calculation is non-trivial
 and has a non-zero answer is somewhat analogous to holographic conformal
 anomaly.

To analyse the structure of the simplified stress tensor, we first need to
recall a few useful results concerning symmetric divergence free tensors from
\cite{MMMV, B, Geroch}. A tensor $\theta_{ab}$ is said to admit a scalar
potential $\alpha$ if \be \label{T2ScalarPot} \theta_{ab}[\alpha] = \cD_a \cD_b \alpha - h_{ab}^{0}  \cD^2 \alpha -
 2 \alpha h_{ab}^{0}.
 \ee
The tensor $\theta_{ab} [\alpha]$ is conserved, i.e., $\cD^a
\theta_{ab}[\alpha]=0$.
Moreover, if $\xi^{a}$ is a Killing vector of $h_{ab}^{0}$ then the current
$\theta_{ab}[\alpha] \xi^{b}$ can be expressed as the divergence of an anti-symmetric tensor \be \label{T2pot}
  \theta_{ab}[\alpha] \xi^{b} = \cD^b\left( 2 \xi_{[b} \cD_{a]}\alpha  + \alpha
  \cD_{[b} \xi_{a]}\right).
\ee
As a result the currents of the form $\theta_{ab}[\alpha] \xi^{b}$ do not
contribute to the conserved charge associated with $\xi^{a}$. Similarly, a
tensor $t_{ab}$ is said to admit a  symmetric, transverse tensor potential 
$\gamma_{ab}$ with $\cD^{a} \gamma_{ab} = 0$ if
\be \label{TensorPotential}
 t_{ab}[\gamma_{ab}] = \cD^2 \gamma_{ab} + 2 \mathcal{R}^{0}_{acbd}\gamma^{cd}
 \quad \mbox{where} \quad\mathcal{R}^{0}_{acbd} = h^0_{ab} h^0_{cd} - h^0_{cb}
 h^0_{ad}.
\ee
The tensor
$t_{ab}[\gamma_{ab}]$ is conserved, and for $\xi^{a}$ a Killing vector of
$h^{0}_{ab}$ the current $t_{ab}[\gamma_{ab}]\xi^{b}$ is the divergence of an
anti-symmetric tensor
\begin{equation}
  t_{ab}[\gamma_{ab}]  \xi^{b} =  2\cD^{a} ( \xi^{c}D_{[a} \gamma_{b]c} +
  \gamma_{c[a}D_{b]} \xi^{c}).
\end{equation}
Hence, currents of this form also do not contribute to the conserved charges.

Our strategy  is to write the simplified expression for the stress tensor after
setting $\sigma = 0$ and $h^2_{ab}$ from \eqref{eqsubs} in terms of a scalar and
a tensor potential.
The simplified boundary stress tensor is $ T^1_{ab}\big{|}_{\sigma = 0} =0$, and
\bea 
 & & T^2_{ab}\big{|}_{\sigma = 0} = \frac{1}{2} \o_a \o_b - 2 \o
\o_{ab} + \o_{ab} \Box \o  
+ \o_{(a}\Box \o_{b)} + \o_{abc}\o^{c} - 4 \o_{ac}\o_b{}^{c} 
+\frac{1}{2}\Box \o_a \Box \o_b 
 \nn \\ & & 
- \frac{1}{2} \Box \o \Box \o_{ab} - 2
\o_{abc}\Box \o^{c}  
+ 2 \o_{c(a}\Box \o_{b)}{}^{c} - \o_{ab} \Box \Box \o 
-\frac{1}{2}\o_{abcd}\o^{cd} 
+ \frac{3}{2} \o_{acd}\o_{b}{}^{cd}
\label{simpTomega} \\ & &
 +h^0_{ab}\left[ \frac{1}{2} \o_c \o^c -\o^2 - 2 \o_c \Box \o^{c} 
+ \frac{1}{2} \Box \o^{c} \Box \o_{c} 
- \frac{1}{2}\o_{cd} \Box \o^{cd} 
+\frac{1}{2} \Box \o \Box \Box \o 
- \frac{1}{2}\o_{cde}\o^{cde} \right].
\nn
\eea
Expression \eqref{simpTomega} can be rewritten as 
\bea
T^2_{ab}\big{|}_{\sigma = 0} = 
2 \theta^{(2)}_{ab} 
- \frac{1}{2} t^{(4)}_{ab}
+ \frac{1}{2} \theta^{(4)}_{(1)ab}
+ \frac{1}{4} s^{(4)}_{ab} 
+ \frac{1}{4} \theta^{(6)}_{(1)ab}
+ \frac{1}{4} \theta^{(6)}_{(2)ab}
- \frac{1}{4} t^{(6)}_{(1)ab}
- \frac{1}{8} t^{(6)}_{(2)ab},
\label{simpTomega2}
\eea
where
\be
 \begin{array}{ll}
\theta^{(2)}_{ab}  = \theta_{ab} \left[ \frac{1}{2} \omega^2 \right],\qquad
\qquad  &
t_{ab}^{(4)} = t_{ab} \left[  \theta^{(2)}_{ab} \right], \\
\theta^{(4)}_{(1)ab} = \theta_{ab} \left[ \omega \Box \omega \right], \qquad
\qquad&
\theta^{(4)}_{(2)ab} = \theta_{ab} \left[ \omega_c  \omega^c \right], \\
\theta^{(6)}_{(1)ab} = \theta_{ab} \left[ \omega_c \Box \omega^c \right], \qquad
\qquad&
\theta^{(6)}_{(2)ab} = \theta_{ab} \left[ \Box \omega
\Box \omega \right],\\
t^{(6)}_{(1)ab} = t_{ab} \left[ s^{(4)}_{ab} \right],
\qquad \qquad &
t^{(6)}_{(2)ab} = 
t_{ab} \left[ \theta^{(4)}_{(2)ab} \right], \\
\end{array}
\ee
and finally
 \bea
 s^{(4)}_{ab} = 2 h^{0}_{ab} \omega_{cd}\omega^{cd} - 2 h^0_{ab} \Box \omega\Box
 \omega - 4 \o_{ac}\o_b{}^{c} + 4 h^0_{ab} \o_c \o^c + 4 \o_{ab} \Box \o - 4
 \o_a \o_b.
 \eea
The superscripts, e.g. as $^{(6)}$ in $t^{(6)}_{(1)ab}$, denote the maximum
number of derivatives appearing in the corresponding expressions. The
subscripts, e.g., $_{(1)}$ in $t^{(6)}_{(1)ab}$, are just labels. We immediately see that with the
possible exception of $s^{(4)}_{ab}$, terms in \eqref{simpTomega2} cannot
contribute to the conserved Lorentz charges. As far as we have explored, we find
that the tensor $s^{(4)}_{ab}$ can possibly contribute to the Lorentz charges.
However, this is not a problem. The contribution due to $s^{(4)}_{ab}$ is simply a c-number
due to our boundary conditions;
it only depends on the background structure $\omega$ and is completely
independent of dynamical fields. Hence, it is a constant over our phase space.
The presence of such a term is consistent with the general analysis of \cite{HIM}.

\subsection{Transformation of Lorentz Charges under Translations}
\label{sec:transform}
Having analysed properties of the stress tensor in special cases in the
previous subsection, now let us study the transformation of Lorentz
charges under translations. The idea behind this computation is as follows. As
mentioned in section \ref{sec:BSTprop1} a general (perhaps somewhat formal) expression for Lorentz charges
can be written as
\be
Q[\xi]  = \lim_{\rho \to \infty}
 \int_{\cC_{\rho}} \sqrt{-h_{C_\rho}} T_{ab} \xi_\rho^a n_{\rho}^{b} d^2x.
\ee
The most important quantity in this expression is the boundary
stress tensor $T_{ab}$, which has expansion in powers of $\rho$.
To investigate transformation property of Lorentz charges under translations, we
need to look at how $T_{ab}$ changes under translations. On the unit
hyperboloid, translations are represented by four functions satisfying
\be
\cD_a \cD_b \chi + h^0_{ab} \chi =0.
\ee
Under translations by an amount $\chi$, the function 
$\o$ changes as $\o \rightarrow \o + \chi.$ We wish to know how the expansion of
the boundary stress tensor changes, i.e., we want to know the expansion of
$\Delta_\chi T_{ab}$.
Since we are considering a difference between two stress tensors for fixed value of $\s$,
many terms immediately cancel out. In particular, in
$\Delta_\chi T_{ab}$ the leading term in the expansion starts at order
$\rho^{-1}$. Due to this fact, calculation of $\Delta_\chi Q[\xi]$ is a
relatively straightforward exercise as opposed to $Q[\xi]$. We find
\bea
\Delta_\chi T_{ab} &=& -\frac{1}{8 \pi G \rho} \left( - 3 \chi h^0_{ab} - 3 \chi
\s_{ab} + h^0_{ab} \s_c \chi^c + \s_{abc} \chi^c\right) + \ldots \\
&=& -\frac{1}{\rho} \cD_c \left[(\s_{ab} + \s h^0_{ab})\chi^c\right] + \ldots~.
\eea
Substituting this expression in the definition of Lorentz charges to calculate
the $\Delta_\chi Q[\xi]$, we see that
\bea
\Delta_\chi Q[\xi] &=& \lim_{\rho \to \infty}
 \int_{\cC_{\rho}} \sqrt{-h_{C_\rho}} \Delta_\chi T_{ab} \xi_\rho^a n_\rho^{b}
 d^2x
 \\
& =&
- \int_{\cC} \sqrt{-h^0_{C}} \cD_c \left[(\s_{ab} + \s h^0_{ab})\chi^c\right]
 \xi^a n_{(0)}^{b} d^2x\\
& =&
 \int_{\cC} \sqrt{-h^0_{C}} \cD_c \left[E^1_{ab}\chi^c\right]
 \xi^a n_{(0)}^{b} d^2x. 
\eea
Here $\cC$ denotes a cut of unit hyperboloid, and $\xi^a$ an exact Killing
vector of the unit hyperboloid, and $n_{(0)}^{b}$ the unit normal to the cut
$\cC$.
This last expression is precisely the expected transformation property of the
Lorentz charges under translations \cite{Ashtekar:1978zz, B,  AshRev}. Note that
the fact that we obtain this result is highly non-trivial. In the
expansion of $\Delta_\chi T_{ab}$ all terms linear in $\omega$ cancel out. Once
again, these remarkable cancellations are highly non-trivial test of  our
computations.

\setcounter{equation}{0}
\section{Conclusions and Future Directions}
\label{sec:conclusions}
Let us summarize what we have achieved in this paper. First and foremost,
we have systematically studied the closest one can come to changing the boundary metric in the
asymptotically flat context, while maintaining the group of asymptotic
symmetries to be Poincar\'e.  The result of this analysis is that we can choose
the  supertranslation frame as we like.
 We studied consequences of making choices $\omega \neq 0$. We performed
 this analysis in the covariant phase space approach as well as in the holographic renormalization approach. 
We showed that the covariant phase space is well
defined irrespective of how we choose to fix supertranslations. Furthermore,
we showed that the on-shell action and the leading order boundary stress tensor
are insensitive to the supertranslation frame. The most significant result of
this paper is the construction of the boundary stress tensor at second order. We
carried out this construction in detail, and studied its conservation
properties.
We also observed that although the next to leading order boundary stress tensor depends
on the supertranslation frame, the dependence is of a very special type. It
is such that the transformation of angular momentum under translations continues
to hold as in special relativity.

Let us now comment on the
motivation Ashtekar and Hansen \cite{Ashtekar:1978zz} had for choosing $\omega
=0$. There it was observed that when $\omega \neq
0$, the second order magnetic part of the Weyl tensors fails to be conserved
with respect to the derivative operator compatible with the unit hyperboloid
metric. This is indeed an obstacle if one insists on using the second
order magnetic part of the Weyl tensor to construct Lorentz charges. However,
this obstacle is only an illusion: above we constructed a symmetric and
divergence free  tensor $V_{ab}$ using second order fields. Taking the curl of
$V_{ab}$ one obtains a new symmetric and divergence free tensor $W_{ab}$
\cite{CDV}. The tensor $W_{ab}$ is the natural quantity to use instead of the
second order magnetic part of the Weyl tensor to construct Lorentz charges
following Ashtekar-Hansen when $\omega \neq 0$.

A
natural extension of our work is to calculate the conserved Lorentz charges
\eqref{charge} using our boundary stress tensor 
 with our supertranslated boundary conditions.  Given the general analysis of
 \cite{MM, Sorkin, HIM}, we expect such a construction to go thorough, however,
 the precise details as to how it proceeds are not investigated at this stage. 
 The reason this computation is non-trivial is because the expression
 \eqref{charge2} is somewhat formal. All quantities that enter into this
 expression admit expansions in powers of $\rho$. This makes the analysis of
 Lorentz charges from holographic point of view significantly
 complicated. We will return to this problem
elsewhere. In this regard, the precise significance of equation
\eqref{conserve1} is also not clear at this stage.

Although boundary stress tensor methods are most well studied for
asymptotically AdS and related settings, the success of these and related
methods in other contexts \cite{MV, MM2, Wiseman, Ross3, Ross2,  BdBH, MM1}
motivates further study in the asymptotically flat context. Our work here attempted to fill in this divide
further by extending our previous work \cite{MMV, MMMV, CDV, V}. We also 
highlighted certain similarities and differences with the asymptotically AdS
setting. Further exploration in this direction should provide additional
insights into the still elusive nature of holography for flat space
\cite{Ma, Barnich:2010eb, deBoer:2003vf, Arcioni:2003xx, Alvarez:2004hga,
Barbon:2008ut, Li:2010dr, Compere:2011dx}.

\subsection*{Acknowledgements}
I thank  Glenn Barnich,  Geoffrey Compere, Francois Dehouck for
discussions  and Donald Marolf and Simon Ross for encouragement.  Several of the
calculations presented in this paper are performed using \textit{xAct}
\cite{JMM}, a suite of free packages for doing tensor algebra in \textit{Mathematica}.
These packages are developed by Jos\'{e} M. Mart\'{\i}n-Garc\'{\i}a and collaborators. I am grateful to 
Leo Stein and the \textit{xAct} Internet community for getting me started. I am
 particularly grateful to Teake Nutma for his help and patient
 explanations on \textit{xAct}, 
and
for sharing his ADM splitting code. I also thank Geoffrey Compere for
his careful reading of an earlier draft of the manuscript and for providing
positive feedback.
\appendix
\setcounter{equation}{0}
\section{Asymptotic Equations of Motion}
\label{app:EOM}
The four dimensional metric is
\be
ds^2 = \left(1 + \frac{\s}{\rho}\right)^2d\rho^2 + h_{ab} dx^a dx^b,
\label{metric}
\ee
where the boundary metric $h_{ab}$ admits an expansion in the inverse powers of
$\rho$ as
\be
h_{ab} = \rho^2 h^0_{ab} + \rho h^1_{ab} +  h^2_{ab} + \ldots~.
\ee
The leading order metric $h^0_{ab}$ is the unit metric on three-dimensional
de-Sitter space. For the considerations of the present paper $h^1_{ab}$
is taken to be of the specific form
\be
h^{1}_{ab} = -2 \s h^0_{ab} + 2 \omega_{ab} + 2 \omega h^0_{ab},
\label{h1}
\ee
where
\be
\omega_{ab} = \cD_a \cD_b \omega. 
\ee
Asymptotic spi-supertranslation Killing vector $\xi^\mu_{\omega}$ is related
to $\omega$ as
\be
\xi^{\rho}_\o = \omega(x) + \mathcal{O}(\rho^{-1}), \qquad
\xi^{a}_\o = \frac{1}{\rho} \o^a(x) + \mathcal{O}(\rho^{-2}),
\ee
where $x$ denotes collectively the coordinates on the three
dimensional de Sitter space, and $\omega$ is an arbitrary smooth function of
these coordinates.

 To obtain the asymptotic equations of motion
we perform the radial 3+1 split. The extrinsic curvature of the constant $\rho$
hypersurfaces can be readily calculated. It admits an expansion in inverse
powers of $\rho$ as,
\bea
K_{ab} &=& \frac{1}{2N} \partial_\rho h_{ab}\\
       &=& \rho h^0_{ab} - 2 \s h^0_{ab} + \omega_{ab} + \omega h^0_{ab} +
       \frac{1}{\rho} (2 \s^2 h^0_{ab} - \s \omega_{ab}- \s \omega h^0_{ab} ) +
       \ldots.
\eea
We are now in position to proceed with a study of asymptotic equations
of motion. We first look at the Hamiltonian `constraint.'

\subsubsection*{Hamiltonian Constraint}
In a simplified form the Hamiltonian constraint reads \cite{BS,B}
\be
h^{ab} \partial_\rho K_{ab} - N K_{ab}K^{ab} + D^2 N =0,
\label{Hcons}
\ee
where $D$ denotes the unique torsion-free covariant derivative compatible with
the full boundary metric $h_{ab}$ and $N$ is the lapse function
\be
N = 1 + \frac{\s}{\rho}.
\ee
 Expansion of
\eqref{Hcons} at the zeroth and first order gives the equation of motion for the
mass aspect
\be
(\Box + 3) \sigma = 0, \label{EOMsigma}
\ee
and at the second order gives the equation for the trace of the second order
metric
\be
h^2 = 12 \s^2 + \s_a \s^a + 3 \omega^2 + 2 \omega \Box \omega +
\omega_{ab}\omega^{ab} - 9 \omega \s - \s \Box \omega + \s_a \omega^a + \s_a
\Box \omega^a + 2 \s_{ab}\omega^{ab}. \label{trace}
\ee
In writing these equations we use the following compact 
notation,
\bea
\o_{abcd} &=& \cD_d \cD_c \cD_b \cD_a \o, \\
\Box \o_{abc} &=& (\cD^e \cD_e) \o_{abc} = \cD^e \cD_e \cD_c \cD_b \cD_a \o,
\eea
and similarly for $\s$. In the following we will use equations \eqref{EOMsigma}
and \eqref{trace} to simplify the resulting expressions.

\subsubsection*{Diffeomorphism Constraints}
In a simplified form the  diffeomorphism constraints read \cite{BS,B}
\be
D_b K^b{}_{a} - D_a K = 0.
\ee
Expansion of this equation at the zeroth and the first orders give no further
non-trivial equation. At the second order it gives the equation for the
divergence of the second order metric
\bea
\cD^b h^2_{ab} &=& 16 \s \s_a + 2 \s_{ab}\s^b + 2 \omega \omega_a + 2 \omega
\Box \omega_a + 2 \omega^b \omega_{ab} + \omega_{ab}\Box \omega^b +
\omega_{abc}\omega^{bc} \nn \\
& & - \s \omega_a - 3 \omega \s_a + \s_a \Box \omega - \s \Box \omega_a + 3
\s_{ab} \omega^b - \o_{ab} \s^b + \s_{ab}\Box \o^b \nn + \s^b \Box \o_{ab} \\ 
&& 
+ 2 \s_{abc}\o^{bc} + 2 \o_{abc} \s^{bc}.
\eea
In the following we will also use this equation in the resulting
expressions.
\subsubsection*{Equations of Motion}
In a simplified form the equations of motion for the boundary metric $h_{ab}$
take the form \cite{BS,B}
\be
F_{ab} := \cR_{ab}- N^{-1} \partial_\rho K_{ab} - N^{-1} D_a D_b N - K K_{ab} +
2 K_{a}{}^{c} K_{cb} = 0.
\ee
These equations can be expanded to give
\be
\cR^0_{ab} = 2 h^0_{ab},
\ee
\be
\cR^1_{ab} = \s_{ab} - 3 \s h^0_{ab} + 3 \omega_{ab} - h^0_{ab} \Box \omega, 
\label{cR1:1}
\ee
\bea
\cR^2_{ab} &=& 2 h^2_{ab} - h^2 h^0_{ab}  + 2 \s_a \s_b - \s
\s_{ab} + (3 \s^2 - \s_c \s^c) h^0_{ab} \nn \\
&&  +  \  (\omega^2 + 3 \omega \Box \omega + 2 \omega_{cd}\omega^{cd})h^0_{ab}
- 7 \omega \omega_{ab} - \omega_{ab} \Box \omega  -2 \omega_{ac}\omega^c{}_{b} \nn
\\ &&
- \  (\s \omega + \s \Box \omega + \s_c\omega^c)h^0_{ab} + 2 \s \omega_{ab} -
\s^c\omega_{abc}.	
\label{cR2:1}
\eea
In expressions \eqref{cR1:1} and \eqref{cR2:1} the left hand side denote the
expansion of the boundary Ricci tensor: 
\be
\cR_{ab} = \cR^0_{ab} + \rho^{-1}\cR^1_{ab}+ \rho^{-2}\cR^2_{ab}+ \ldots~.
\ee
In order to obtain the asymptotic equations of motion in the most useful form we
now express $\cR^1_{ab}$ and $\cR^2_{ab}$ as appropriate derivative operators acting
on the metric components. The explicit forms for
$\cR^1_{ab}$ is 
\bea
\cR^1_{ab} &=& \cD_c \cD_{(a}h^1_{b)}{}^{c}- \frac{1}{2} \cD_a \cD_b h^{1}-
\frac{1}{2} \Box h^{1}_{ab},
\label{cR1:2}
\eea
and similarly for $\cR^2_{ab}$ is
\bea
\cR^2_{ab} &=& \cD_c \cD_{(a}h^2_{b)}{}^{c}- \frac{1}{2} \cD_a \cD_b h^{2}-
\frac{1}{2} \Box h^{2}_{ab} + \frac{1}{2}h^{1cd} \cD_a \cD_b h^1_{cd} -
h^{1cd}\cD_c \cD_{(a} h^1_{b)d} \nn \\ &&  
+ \frac{1}{4}\cD_a h^1_{cd} \cD_b h^{1cd} + \cD^dh^1_b{}^{c} \cD_{[d} h^1_{c]a}
+ \frac{1}{2} \cD_d(h^{1cd} \cD_c h^1_{ab}) - \frac{1}{4}\cD^c h^1 \cD_c
h^1_{ab} \nn  \\ & & - \left(\cD_d h^1{}^{cd} - \frac{1}{2} \cD^c
h^1\right)\cD_{(a}h^1_{b)c} \label{EOM1:2}.
\eea
These expressions can be expanded to give respectively,
\bea
\cR^1_{ab} &=& \s_{ab} + h^0_{ab} \Box \sigma + 3 \omega_{ab} - h^0_{ab} \Box
\omega, \\
\cR^2_{ab} &=&  \cD_c \cD_{(a}h^2_{b)}{}^{c}- \frac{1}{2} \cD_a \cD_b h^{2}-
\frac{1}{2} \Box h^{2}_{ab} + 3 \s_a \s_b + 2 \s \s_{ab} + 2 h^0_{ab} \s \Box \s
+ h^0_{ab} \s_c \s^c \nn \\
& & - 6 \omega \omega_{ab} 
- 2 \omega_{ab} \Box \omega 
+(2\omega \Box \omega 
- \omega^c \Box \omega_c   
+ 2 \omega_{cd}\omega^{cd}) h^0_{ab} 
+ \omega_{abc}\omega^c 
-\omega_{abc}\Box \omega^c
+\omega_{acd}\omega_{b}{}^{cd} \nn \\
& & + 6 \s \omega_{ab} - 2 \omega \s_{ab} 
- 2 \s_{ab} \Box \omega 
- (2  \s \Box \omega 
+ 2  \omega \Box \s 
+ 2  \s_c \omega^c 
+    \s^c \Box \omega_c 
+ 2  \s_{cd}\omega^{cd})h^0_{ab} \nn \\
& & 
- \omega_{abc}\s^c  
+ 4 \s_{(a}{}^{c} \omega_{b)c}.
\eea
Equating \eqref{cR1:1} and \eqref{cR1:2} gives the equations of motion for the
first order metric components.  We obtain again
$\Box \s + 3 \s = 0.
$ In particular, upon equating \eqref{cR1:1} and \eqref{EOM1:2} we do not obtain
any new non-trivial equation. At the second order we do obtain a non-trivial equation---the equation of motion for
$h^2_{ab}$. After a significant amount of algebra it reads
\bea
(\Box - 2 )h^2_{ab} &=& 	6(\s_c \s^c - 3 \s^2)h^0_{ab} + 8 \s_a \s_b + 14 \s
\s_{ab} + 2 \s_{ac}\s^{c}{}_{b} + 2 \s_{abc}\s^c\nn \\
& & + 2  (\omega \Box \o - \o^2 + \o_c \o^c )h^0_{ab} 
 - 4 \o \o_{ab} + 2 \o_{ab}\Box \o + 2 \o \Box \o_{ab}
 \nn \\ &&
  + 4 \o_{abc}\o^c
-2 \o_{cb}\o^{c}{}_{a} + 2 \o_{a}{}^{cd} \o_{bcd} + 2 \o_{c(a} \Box \o_{b)}{}^c
\nn \\
& & + (14 \o \s - 4 \s \Box \o - 4 \s_c \o^c + 2 \s^c \Box \o_c + 4
\s_{cd}\o^{cd})h^0_{ab} + 17 \s \o_{ab} - \o \s_{ab} \nn \\
& & 
- \s_{ab}\Box \o - \s \Box \o_{ab} + 5 \s_{abc}\o^c - 5 \s^c \o_{abc} + \s_{abc}
\Box \o^c + \s^c \Box \o_{abc} \nn \\
& &  + 2 \s_{abcd}\o^{cd} + 2 \o_{abcd}\s^{cd} + 2 \s_{c(a} \o_{b)}{}^c  + 2
\s_{c(a} \Box \o_{b)}{}^c + 4 \s_{(a}{}^{cd} \o_{b)cd}. 
\eea
In writing all second order  equations above we have carefully separated 
$(\s,\s)$, $(\o,\o)$, and $(\s,\o)$ terms.

\setcounter{equation}{0}
\section{Certain Details on the Boundary Stress Tensor Computation}
\label{app:BST}
This appendix contains certain details on the boundary stress tensor
computation. We make use of the asymptotic equations of motion as needed. The
calculation is organized as in appendix B of \cite{MMMV}. We begin by
calculating the expansion of the tensor $\tilde L^{ab}$.  It is defined as
\be
\tilde L^{ab} = h^{cd} (L^{-1})_{cd}{}^{ab},
\ee
where
\be
L_{ab}{}^{cd} = h^{cd}\hat K_{ab} + \delta_{(a}^{(c}\delta_{b)}^{d)} \hat K -
\delta_{(a}^{(c}\hat K_{b)}^{d)}- \delta_{(a}^{(d}\hat K_{b)}^{c)},
\ee
and 
\be
(L^{-1})_{cd}{}^{ab}(L)_{ab}{}^{ef} = \delta_{(c}^e \delta_{d)}^f.
\ee
After a straightforward, but tedious, computation we find the expansion of 
$\tilde L^{ab}$ in the inverse powers of $\rho$ to be as follows,
\bea
\tilde L^{ab} &=& \frac{1}{4 \rho}h^{0 ab} + \frac{1}{\rho^2}\left[
\frac{1}{2}\s^{ab} + h^{0ab} \s - \o^{ab} + \frac{1}{4}(\Box \o -
\o)h^{0ab}\right] + \frac{1}{\rho^3} 
\Bigg{\{}\left. 
- \frac{1}{4}h^{2ab}  
\right. 
\nn \\ & & 
\left. + \frac{99}{16}h^{0ab} \s^2
- \frac{1}{4} h^{0ab} \s_c \s^c + \frac{3}{2} \s^{ac}\s_{c}^{b} -
\frac{9}{16}h^{0ab}\s_{cd}\s^{cd} + \s^a \s^b + \frac{9}{2}\s\s^{ab} 
 \right. 
 \nn \\ & &
 \left.  
+ \frac{7}{2} \o^{a}_{c} \o^{cb}
- \frac{3}{2} \o^{ab}\Box \o  
+ \frac{5}{2}\o^{ab} \o 
+ \frac{1}{2} \left[\o^2 - \o \Box \o +
\frac{3}{4} (\Box \o)^2 - \frac{5}{4} \o_{cd}\o^{cd}\right] h^{0ab} 
\right. 
\nn \\ & & 
\left.
-8 \s \o^{ab} - \frac{1}{2} \o \s^{ab} + \frac{3}{2} \Box \o \s^{ab} - 5
\s_c{}^{(a} \o^{b)c} - \frac{1}{2} \o^{abc}\s_c + \left[- 2 \o \s +
\frac{29}{8} \s \Box \o 
\right. \right. 
\nn \\ &&
\left. \left.
- \frac{1}{4} \s_c \o^c + \frac{1}{4}
\s_c \Box \o^c + \frac{13}{8} \s_{cd}\o^{cd}\right] h^{0ab} 
\right. \Bigg{\}}+
{\mathcal O} \left(\frac{1}{\rho^{4}}\right)~.
\eea
Given this expression and the expansion of $\hat K_{ab}$, it is 
straightforward to compute the expansion of the holographic stress tensor
$T_{ab}$
\be
T_{ab} = - \frac{1}{8\pi G}\left(\pi_{ab} - \hat \pi_{ab} + \Delta_{ab}\right),
\ee
where
\be
\pi_{ab} =  K h_{ab}-  K_{ab} \quad \mbox{and} \quad \hat \pi_{ab} = \hat K
h_{ab}- \hat K_{ab},
\ee
and where $\Delta_{ab}$ is
\be
\Delta_{ab} = \hat K_{ab} - 2 \tilde L^{cd} (\hat K_{cd} \hat K_{ab} - \hat
K_{ac} \hat K_{db}) + D^2 \tilde L_{ab} + h_{ab}D_c D_d \tilde L^{cd} - 2
D_{d}D_{(a}\tilde L_{b)}{}^{d}.
\label{Delta}
\ee
A straightforward computation gives
\bea
\pi_{ab} - \hat \pi_{ab} &=& \s_{ab} + \s h^0_{ab} + \frac{1}{\rho}\bigg{[}
h^2_{ab} + 2 \s_a \s_b + \s \s_{ab}  - \left(\frac{5}{2}
\s^2 + \s_c \s^c + \frac{1}{2}\s_{cd} \s^{cd} \right)h^0_{ab} \nn \\
&& + \s_{ac}\s^{c}{}_{b} - h^0_{ab} \o^2 - 2 \o \o_{ab} -
\o_{ac}\o^{c}{}_{b} + \s \o_{ab} + \o \s_{ab} + \s_{ab} \Box \o -2
\s_{c(a}\o^c{}_{b)} \nn \\ &&  
- \o_{abc} \s^c + (4 \s \o + 2 \s \Box \o - \s_c \o^c + \s_{cd}\o^{cd})h^0_{ab}
\bigg{]} + \ldots~.
\eea
Similarly,
\bea
 \hat K_{ab} - 2 \tilde L^{cd} (\hat K_{cd} \hat K_{ab} - \hat
K_{ac} \hat K_{db}) &=&\frac{1}{2} \left( 3 \s_{ab} + 3 \s
h^0_{ab} -
3\o_{ab} + \Box \o h^0_{ab} \right) 
+ \frac{1}{\rho} \bigg{[} 3 \s_a \s_b 
\nn \\ && \hspace{-1.8 in} 
+  \frac{21}{2} \s \s_{ab} + 6
\s_{ac}\s^{c}{}_{b} + \left(\frac{21}{2}\s^2 - \s_c \s^c - 2
\s_{cd}\s^{cd}\right)h^0_{ab} - 2 \o_{ab} \Box \o + 3 \o_{ac}\o^{c}{}_{b} 
\nn \\ && \hspace{-1.8 in} 
+ h^0_{ab} (\Box \o)^2 - 2 h^0_{ab} \o_{cd}\o^{cd} - 9 \s \o_{ab} + \frac{3}{2}
\o \s_{ab} + \frac{9}{2} \s_{ab} \Box \o - \frac{3}{2}\o_{abc}\s^c - 12
\s_{c(a}\o_{b)}{}^{c} 
\nn \\ && \hspace{-1.8 in} 
+ \left(\frac{3}{2}\s \o + \frac{17}{2} \s \Box \o - \s_c \o^c +
\frac{1}{2}\s_c \Box \o^c + 5 \s_{cd}\o^{cd}\right)h^0_{ab} \bigg{]} + \ldots~.
\eea
A calculation of the derivative terms requires more work. We find
\bea
D^2 \tilde L_{ab} + h_{ab}D_c D_d \tilde L^{cd} - 2 D_{d}D_{(a}\tilde
L_{b)}{}^{d} &=& - \frac{3}{2} \s_{ab} - \frac{3}{2}\s h^0_{ab} + \frac{3}{2}
\o_{ab} - \frac{1}{2}h^0_{ab} \Box \o +  \frac{1}{\rho} \Bigg{\{} -3 \s_a \s_b
\nn \\ && \hspace{-1.9 in} 
+\frac{3}{4} \s^2 h^0_{ab} + \frac{3}{4}\s \s_{ab} + 5 h^0_{ab}\s^c \s_c +  4
\s_{abc}\s^c - \frac{3}{4} \s_{cd}\s^{cd} h^0_{ab} - \frac{3}{4}\s_{abcd}\s^{cd}
+
\frac{9}{4}\s_{acd}\s_b{}^{cd} 
\nn \\ && \hspace{-1.9 in}
 - \frac{3}{4} h^0_{ab} \s_{cde}\s^{cde} +  \bigg{[} \frac{1}{2}\o_c
 \o^c - \Box \o \Box \o - 2 \o^c \Box \o_c + 2 \o_{cd}\o^{cd} 
 + \frac{1}{2} \Box
 \o_c \Box \o^c
+ \frac{1}{2}\Box \o \Box \Box \o
 \nn \\ && \hspace{-1.9 in}
  - \frac{1}{2} \o^{cd}\Box \o_{cd} 
  - \frac{1}{2} \o_{cde}\o^{cde} \bigg{]} h^0_{ab} + \frac{1}{2} \o_a \o_b  + 3
  \o_{ab} \Box \o + \o_{(a}\Box \o_{b)} +  \o_{abc}\o^c - 6 \o_{ac}\o_{b}{}^{c} 
\nn \\ && \hspace{-1.9 in}   
  + \frac{1}{2} \Box \o_a \Box \o_b ~ - ~ \frac{1}{2}\Box \o \Box \o_{ab} ~ -
  ~ 2 \o_{abc}\Box \o^c ~ + ~  2 \o_{c(a} \Box \o_{b)}{}^{c} -  \o_{ab}\Box
  \Box \o  ~ - 
  \frac{1}{2}\o_{abcd}\o^{cd} 
\nn \\ && \hspace{-1.9 in}   
+ \frac{3}{2} \o_{acd}\o_{b}{}^{cd} ~ + ~9 \s \o_{ab}- \frac{3}{2}\o \s_{ab} ~ -
~
\frac{15}{4} \s_{ab}\Box \o ~ - ~ \frac{9}{4} \s \Box \o_{ab} ~ - ~ \frac{3}{2}
\s_{abc}\o^{c}  -  3 \o_{abc}\s^c 
\nn \\ && \hspace{-1.9 in}   
+ 12 \s_{c(a}\o_{b)}{}^{c} 
~+ 3 \Box \o_c \s_{ab}{}^{c} 
~- 3 \s_{c(a}\Box \o_{b)}{}^{c} 
~+ \frac{3}{2} \s_{ab} \Box \Box \o 
~+ \frac{3}{4} \s_{abcd}\o^{cd}
~+ \frac{3}{4} \o_{abcd}\s^{cd} 
\nn \\ && \hspace{-1.9 in}   
- \frac{9}{2} \s_{(a}{}^{cd}\o_{b)cd} 
+ h^0_{ab} \bigg{[} 4 \s^{c}\Box \o_c 
~- \frac{3}{2} \s \o 
~- \frac{25}{4} \s \Box \o 
- \frac{7}{2} \s_c \o^c   
+ \frac{9}{4} \s \Box \Box \o 
- \frac{11}{4} \s_{cd}\o^{cd} 
\nn \\ && \hspace{-1.9 in}   
+ \frac{3}{4}
\s_{cd}\Box\o^{cd} +  \frac{3}{2} \s_{cde}\o^{cde}  
\bigg{]} \Bigg{\}} + \ldots~.
\eea
Putting all this together we obtain a final expression for the boundary
stress tensor to the relevant order. Such an expression is presented in the main
text \eqref{mainT}.

\end{document}